\documentclass[%
 aip,
 amsmath,amssymb,
preprint,
]{revtex4-1}

\usepackage{graphicx}
\usepackage{dcolumn}
\usepackage{bm}

\usepackage[utf8]{inputenc}
\usepackage[T1]{fontenc}
\usepackage{mathptmx}
\usepackage{etoolbox}

\usepackage{amsmath}
\usepackage{amssymb}

\makeatletter
\def\@email#1#2{%
 \endgroup
 \patchcmd{\titleblock@produce}
  {\frontmatter@RRAPformat}
  {\frontmatter@RRAPformat{\produce@RRAP{*#1\href{mailto:#2}{#2}}}\frontmatter@RRAPformat}
  {}{}
}%
\makeatother
\begin{document}


\title[Effect of bismuth crystal orientations in Nernst thermomagnetic devices]{Effect of bismuth crystal orientations in Nernst thermomagnetic devices}
\author{A. Sola}
 \email{a.sola@inrim.it}
\author{E. S. Olivetti}%
\author{A. Di Pietro}%
\author{L. Martino}%

\author{V. Basso}
\affiliation{%
Istituto Nazionale di Ricerca Metrologica, Strada delle Cacce 91, 10135 Turin, Italy
}%

\date{\today}

\begin{abstract}
In this work we report Nernst effect measurements in single crystal bismuth samples, with special emphasis on the characterization of the Nernst coefficient when the magnetic field, heat current and generated voltage are aligned along specific directions relative to the crystal axes. We found significant differences between the different orientations, reflecting the highly anisotropic electronic structure of bismuth and compatible with the Nernst characteristics obtained from polycrystalline samples.
These results not only complement the experimental works published in the past but also underline the role of crystalline orientation in the context of transverse thermoelectric effects, towards an efficient design of thermomagnetic devices like the ordinary-Nernst-effect-based energy harvesters.

\end{abstract}
\maketitle

Thermoelectricity is of great interest both for the study of fundamental phenomena and for technological applications \cite{nolas2001thermoelectrics,mao2021thermoelectric}. 
When the magnetic degree of freedom is taken into account, thermoelectric effects are called thermomagnetic or Nernst-Ettingshausen effects\cite{nolas2001thermoelectrics} and they exhibit a transverse geometric configuration. The technological interest on these topics is driven by the opportunity to use a single material instead of junctions, resulting in a greater freedom in the design of devices\cite{boona2021transverse,uchida2022thermoelectrics}.
In a solid that exhibits transverse thermoelectricity the generated voltage, the magnetic field and the heat current are perpendicular to each other.
It is possible to observe thermomagnetic effects both in magnetic materials \cite{uchida2021transverse} and in normal metals or semimetals \cite{akhanda2024thermomagnetic} under magnetic field. In the first case, the thermomagnetic effect is also called anomalous Nernst effect\cite{mizuguchi2019energy} in analogy to the anomalous Hall effects. The advantage, with respect to the ordinary effect, is the absence of an external magnetic field. However, according to the literature\cite{sakai2018giant}, the materials with the best anomalous Nernst coefficient have lower figure of merit values than the ones reported for the ordinary Nernst effect of bismuth and bismuth-antimony alloys. 
The attention to the ordinary Nernst effect in the field of thermoelectric conversion had already been highlighted in the past\cite{angrist1963nernst}, especially with regards to bismuth and bismuth-antimony\cite{harman1964nernst}. More recently, the transverse orientation of the ordinary Nernst effect has been exploited for the design of non-flat thermomagnetic devices\cite{de2016thermomagnetic,yang2017scalable}. An ordinary-Nernst-effect-based transverse thermoelectric module equipped with embedded permanent magnets has been recently reported by Murata et al.\cite{murata2024zero}, with the advantage of the large ordinary Nernst thermopower of bismuth-antimony, that, together with the one of elemental bismuth, is the largest reported in the literature\cite{harman1964nernst}.

Elemental bismuth has been the subject of research for decades: its peculiar electronics structure \cite{yang2000shubnikov} gives rise to characteristic properties which have aroused curiosity of scientists since the beginning of research on solid state physics \cite{fuseya2015transport}, including the discovery of the Seebeck and Nernst effects \cite{seebeck1895magnetische,v1886ueber}.
Despite the vast literature on the thermoelectric effects of bismuth, a comprehensive experimental characterization of its low-field Nernst behaviour as a function of crystal orientation is still missing. Our study is motivated by the need to have detailed experimental data in order to design ordinary-Nernst-effect-based transverse thermoelectric devices.

\begin{figure}
\includegraphics[width=1\linewidth]{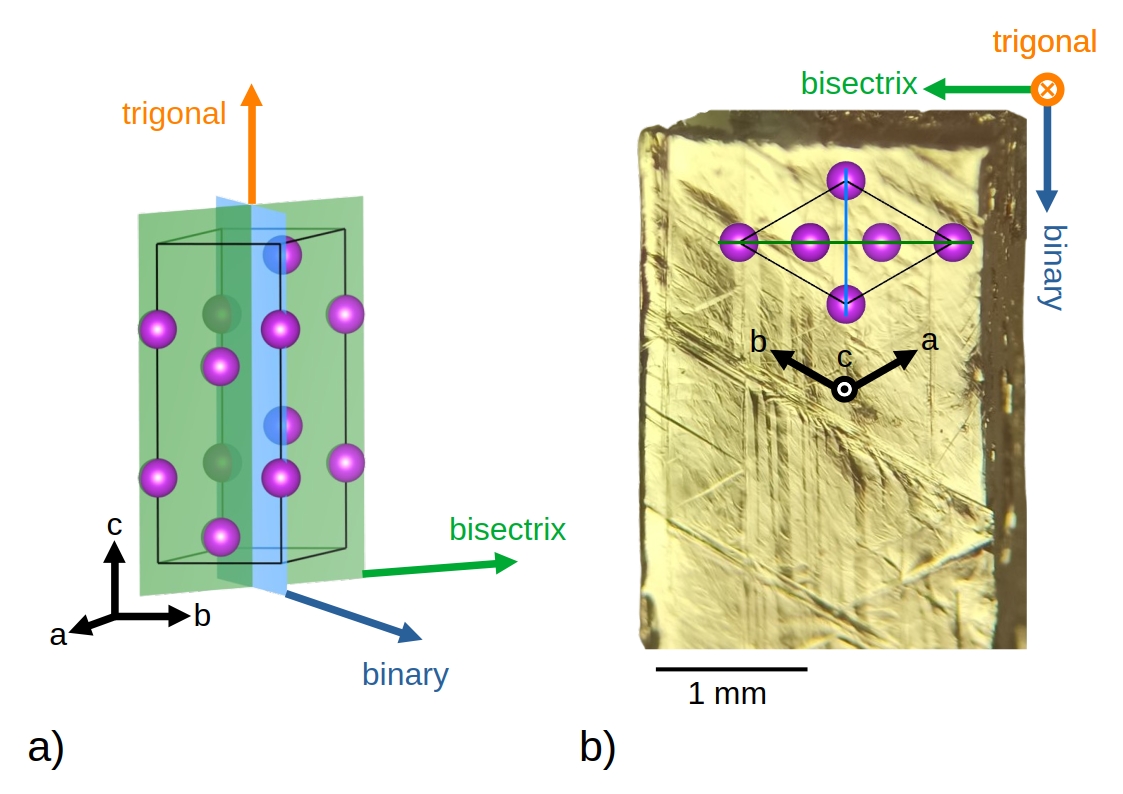}
\caption{\label{Fig1} a) Perspective view of the unit cell of bismuth represented with hexagonal axes; b) Projection of the unit cell along the [001] direction superimposed to an optical microscope image of the sample, which exposes the plane perpendicular to the trigonal axis (basal plane). The directions of the binary and bisectrix axes are also reported.}
\end{figure}

\begin{figure}
    \centering
    \includegraphics[width=1\linewidth]{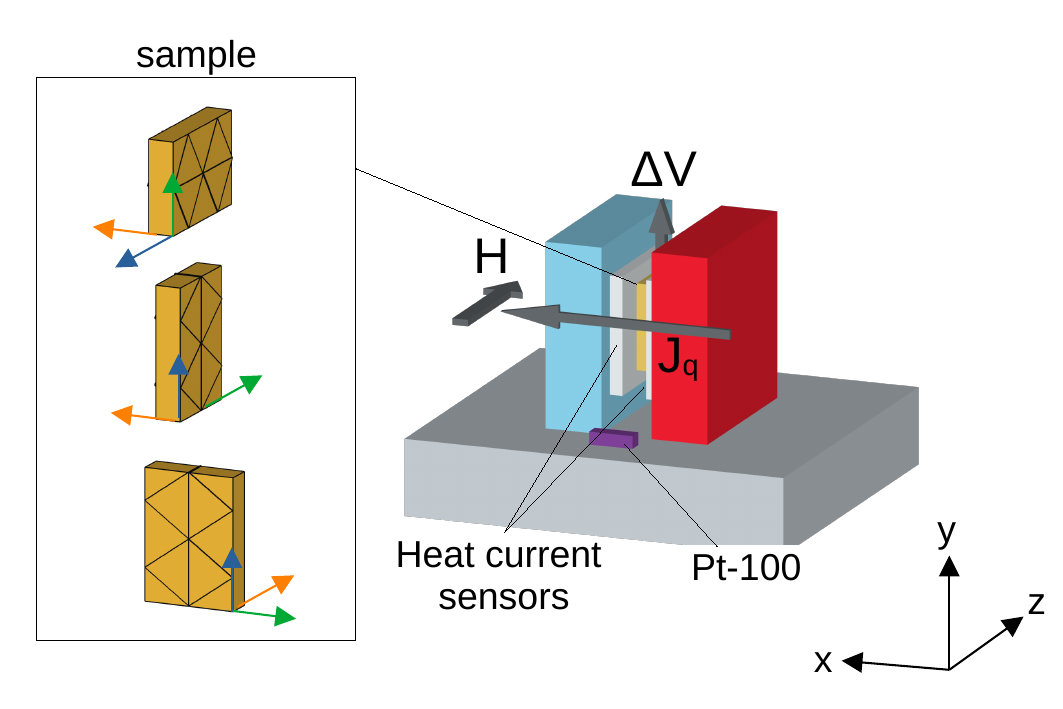}
    \caption{On the left, the bismuth crystal is shown with the orientations relevant for the measurements. This is positioned inside the measurement system, represented on the right. The blue, green and orange arrows on the sample represent the binary, the bisectrix and the trigonal axes, respectively.}
    \label{Fig2}
\end{figure}

We prepared both polycrystalline and single-crystal bismuth samples. The polycrystalline sample is obtained by pressing bismuth powders into a die of parallelepiped shape (6.95 x 4.60 x 0.8 mm$^3$). The single-crystal samples were obtained by cleaving and cutting a parallelepiped (6.85 x 2.85 x 0.6 mm$^3$) from a larger chunk of 99.999\% bismuth with cm grain size. The pieces were easily cleaved with a sharp blade along the basal plane which is the primary cleavage plane of Bi\cite{brown1968tilt}. The freshly cleaved surface has terraces whose orientation can be exploited to locate the binary axes, and, by consequence, the bisectrix direction perpendicular to it, in order to cut the edges of the samples accordingly. The out of plane orientation of the c-axis of the selected crystals was confirmed by X-ray diffraction, which revealed a strong (00n) texture on both surfaces of the sample; a secondary (n0n) orientation was also observed, with reflections of much smaller intensity, revealing that a minor portion of the sample is constituted by grains or domains with the {n0n} planes parallel to the surface. We believe this can be due to crystal twinning induced by mechanical deformation occurred during cutting, since it was reproducibly observed in all the samples cleaved from larger crystals. A magnified image of the single crystal sample is reported in Fig.\ref{Fig1} together with a sketch of the rhombohedral unit cell of bismuth represented with hexagonal axes.
For what concerns the measurement system, we employed an experimental setup, previously described \cite{sola2017longitudinal,venkat2020measurement}, suitable for the quantitative characterization of transverse thermoelectric effects such as the Nernst effect \cite{sola2023polycrystalline} and the spin Seebeck effect \cite{sola2018spincaloritronic}. A scheme of the setup is shown in Fig.\ref{Fig2}, together with the geometrical configurations of the measurements on the bismuth single crystal.
The characterization of thermomagnetic effects requires the measurement of three quantities: the magnetic field, the generated voltage and the thermal gradient. The first one is obtained by an electromagnet and a Hall probe, the second one is a standard DC voltage measurement and the third one is based on the heat flux method, which consists of a procedure aimed at avoiding the effects of the thermal resistance of the contacts between the temperature sensors and the sample under test. This is possible because the thermoelectric response of the sample is driven by a controlled heat current through it. This heat current is measured by a pair of calibrated Peltier sensors connected to two parallel faces of the test sample. The control of heat current is further ensured by setting the experiment under a vacuum of $10^{-7}$ mbar and by electrically connecting the sample with wires of reduced cross section (0.012 mm).
The sensitivity $K$ of the Peltier sensors is determined according to a calibration procedure reported in a previous work\cite{venkat2020measurement} and the value of $K$ that we found in the present work is represented by the following expression:
\begin{equation}
K = 0.88 + 3.86\cdot10^{-3}T-1.39\cdot10^{-5}T^{2}-0.93\cdot10^{-7}T^{3}
\tag{1}\label{eq:1}
\end{equation}
as a function of the temperature $T$ in °C at which the experiment is conducted. This temperature is the one of the whole system formed by sensors, thermal baths and sample under test when the experiment is at equilibrium and it is measured by a standard Pt-100 sensor placed on the thermal reservoir at which the experiment is connected, as shown in Fig.\ref{Fig2}. A controller is able to set this temperature in the range between 230 K and 350 K.
This experimental method allows us to evaluate the thermopower $N$ as
\begin{equation}
N=\frac{\nabla_{\mathrm{y}}V}{\nabla_{\mathrm{x}}T}=\frac{V_{\mathrm{NE}}/L_{\mathrm{y}}}{-j_{\mathrm{q}}/k} \tag{2}\label{eq:2}
\end{equation}
where $V_{\mathrm{NE}}$ is the Nernst voltage generated by the sample, $j_{\mathrm{q}}$ is the heat current density through it, $L_{\mathrm{y}}$ is the distance between the electrodes connected on the sample and $k$ is the thermal conductivity of the material.
\begin{figure}
    \centering
    \includegraphics[width=1\linewidth]{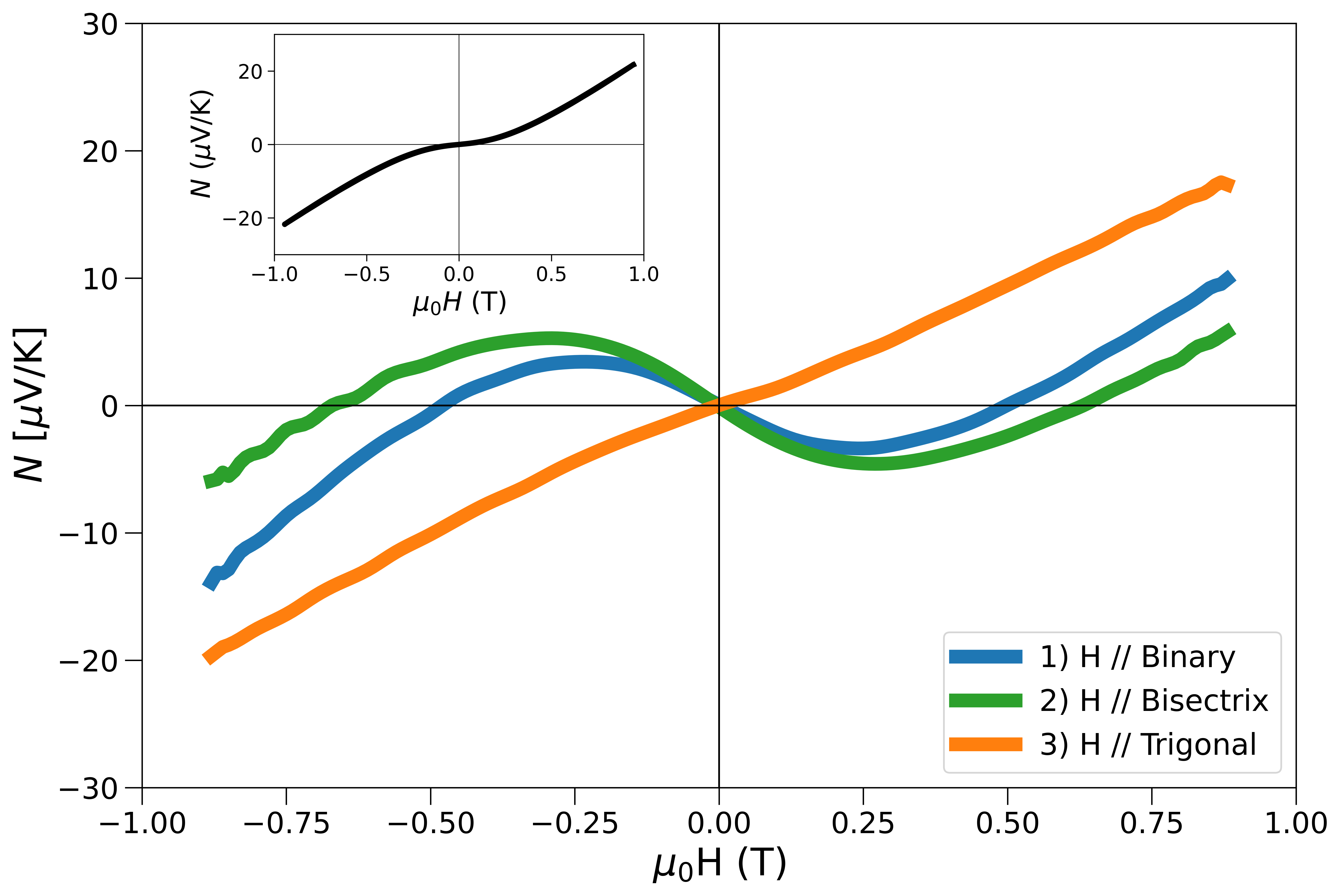}
    \caption{Nernst thermopower vs. magnetic field characteristics of the crystal bismuth, with magnetic field along binary (blue line), bisectrix (green line) and trigonal (orange line) axes. Inset: characteristics of a polycrystalline sample.}
    \label{Fig3}
\end{figure}
The ordinary Nernst effect characterization provides the thermopower $N$ as a function of the magnetic field $H_{\mathrm{z}}$. The quantitative evaluation of $N$ as well as the comparison of experimental results at different conditions of the sample under test is ensured by the heat flux method. This procedure requires as input parameter the value of thermal conductivity $k$, since $N$ is represented by the ratio between the electric and the thermal gradients. Usually the value of thermal conductivity $k$ is well known, especially for elemental crystals; nevertheless in the case of crystalline bismuth it is worth taking anisotropies into account. In particular, the literature reports a factor of $1.39$ between the values of thermal conductivity measured along and perpendicular to the trigonal axis, respectively\cite{kaye1923thermal}.
For what concerns the application of a magnetic field, with the heat current perpendicular and the magnetic field parallel to the trigonal axis a decrease of 16\% of the value of thermal conductivity has been reported: this scales down to 7.3\%, with a 90 degree rotation\cite{kaye1929cviii}.
The anisotropy of electrical conductivity, whose value affects the thermomagnetic figure of merit, has been discussed in the past \cite{kaye1939thermal}: the values of electrical resistivity along the directions perpendicular and parallel to the trigonal axis are 1.14$\cdot 10^{-6}$ $\Omega \cdot$m and 1.44$\cdot 10^{-6}$ $\Omega \cdot$m, respectively. The magnetoresistance of bismuth along each crystal orientation is also reported by the same authors \cite{kaye1939thermal}.

The results of the Nernst thermopower vs. magnetic field characteristics, obtained with the crystal orientations sketched in Fig.\ref{Fig2}, are reported in Fig.\ref{Fig3}.
The values obtained with the magnetic field along the binary axes, represented by the blue line in Fig.\ref{Fig3}, and with the magnetic field along the bisectrix axes, represented by the green line in Fig.\ref{Fig3}, are in agreement with previous works\cite{michenaud1971field,hansen1977weak}. 
The slight asymmetry between the positive and the negative branches of the curves that is observed for the magnetic field along binary and bisectrix axes is probably due to a slight uncontrolled misalignment of the experiment that causes the rising of a spurious magneto-Seebeck component \cite{spathelf2022magneto}. Instead the presence of Umkehr effect can be excluded at room temperature \cite{lenoir2002umkehr}.
In the inset of Fig.\ref{Fig3} is shown the Nernst thermopower vs. magnetic field characteristics that we measured on a polycrystalline sample. This exhibits a low-field deviation from linearity that has been previously observed and reported in the literature by other authors \cite{de1992nernst,hamabe2003magnetic,zhang2015control,parzer2024measurement}.
\begin{figure}
    \centering
    \includegraphics[scale=0.32]{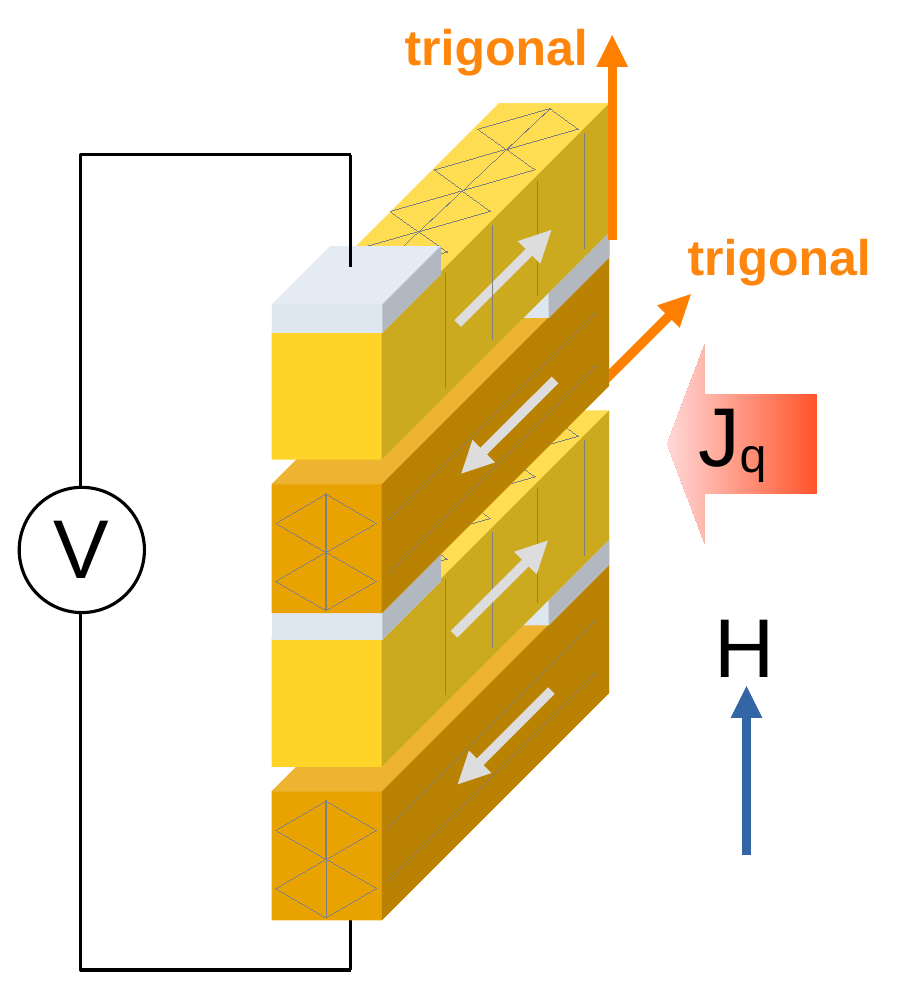}
    \caption{Possible design of a Nernst thermopile based on single crystal bismuth elements in series, cut with the trigonal axes parallel and perpendicular to the applied magnetic field. The white arrows represent the electric current contributions from each element.}
    \label{Fig5}
\end{figure}
For a further investigation of the behaviour of bismuth single crystals, we repeated the Nernst thermopower vs. magnetic field measurements in a temperature range close to room temperature. The experimental data are shown in Fig.\ref{Fig4} for the same geometries adopted for the previous measurements, with the magnetic field along binary (blue dataset), bisectrix (green dataset) and trigonal axes (orange dataset).
\begin{figure*}
\includegraphics[scale=0.50]{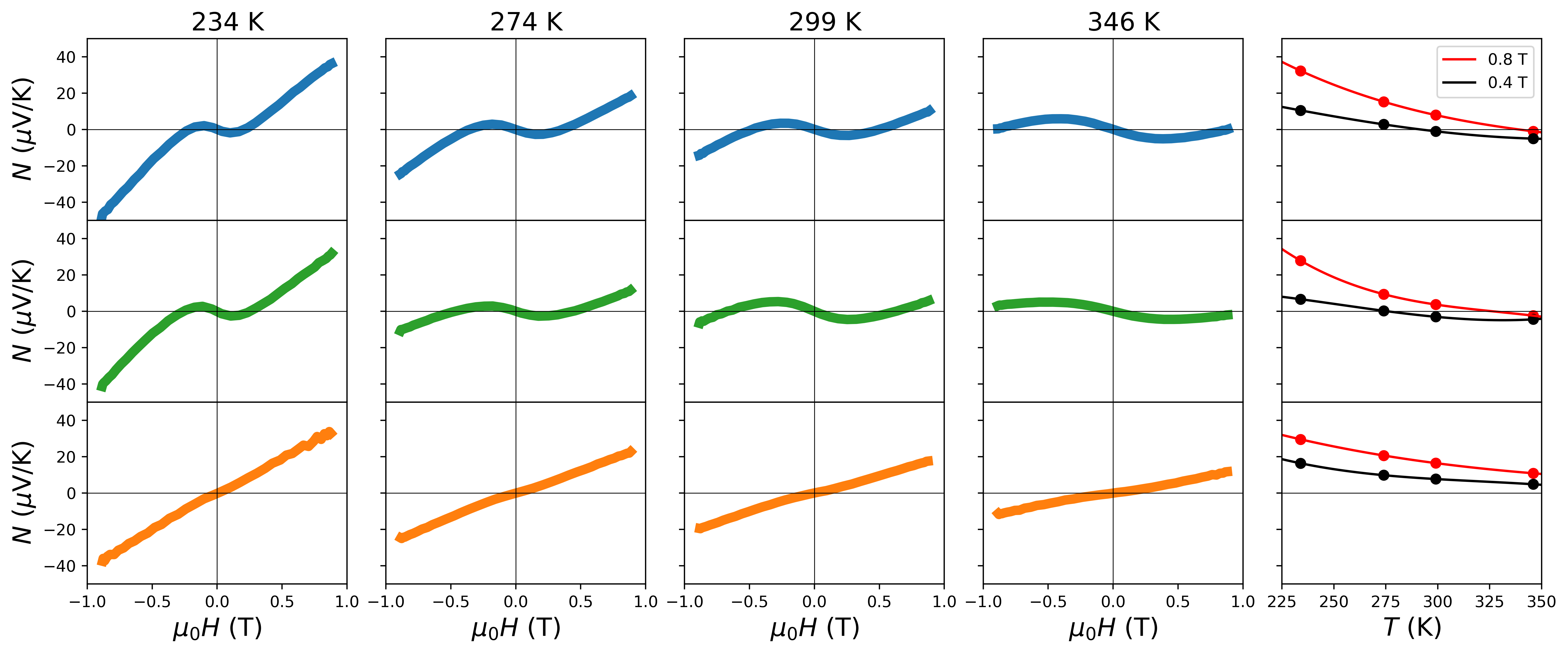}
\caption{\label{Fig4} Nernst thermopower vs. magnetic field characteristics of the crystal bismuth sample at different temperatures. The characteristics shown with blue lines are obtained with magnetic field along binary axes, the ones in green refers to the bisectrix axes and the ones in orange to the trigonal axis. In the last column the temperature dependence of the Nernst thermopowers measured at 0.8 T (red points) and 0.4 T (black points) is reported as a summary.}
\end{figure*}
From the measurements reported in Fig.\ref{Fig4} we can observe that the magnitude of the effect increases as the temperature decreases, as expected for the Nernst effect in bismuth in this temperature range\cite{nolas2001thermoelectrics}. Moreover, the temperature dependent behaviour of the Nernst thermopower with the magnetic field  aligned along the binary axes (blue data of Fig.\ref{Fig4}) is in agreement with data reported in the literature\cite{michenaud1971field} for the temperature range between 195 K and 300 K. Also the temperature dependent displacement of the local maximum of the Nernst thermopower vs. magnetic field curve corresponds to the measurements reported by Michenaud et al.\cite{michenaud1971field}.

In conclusion, we conduced an experimental study that represents a synthesis of the Nernst properties of bismuth at low magnetic fields, between +1 T and -1 T, and in a temperature range between 234 K and 347 K, taking into account the orientation of the single crystal sample with respect to the directions of the magnetic field, the heat flow and the generated electric potential.
We found sign changes in the Nernst thermopower vs. magnetic field characteristics for different orientations of the bismuth crystal. 
The potential applications of this study are the design of optimized thermomagnetic devices in the field of heat sensors or energy harvesters, like the modules based on ordinary Nernst effects \cite{murata2021prototype,murata2024zero} and transverse thermopiles whose working principle is based on Nernst thermopowers of opposite signs, as in the case of FePt and MnGa epitaxial films proposed by Sakuraba et al. \cite{sakuraba2013anomalous}. 
In the case of bismuth, it could be possible to exploit the different signs exhibited by the Nernst thermopower when the magnetic field is applied parallel and perpendicular to the trigonal axis. This could open to the design of a Nernst thermopile assembled by bismuth oriented crystals, as represented in Fig.\ref{Fig5}.
Finally, the detailed investigation of the Nernst characteristic of a single crystal is a clarifying element for the behaviour of polycrystalline samples for what concerns the deviations from linearity that are observed al low fields.

\begin{acknowledgments}
This project has received funding from the Italian ministry of university and research under the Next-Generation Metrology project and under the Research Projects of Relevant National Interest (PRIN project Xverse T.E.C “Transverse thermoelectric energy conversion”: grant no. 2022LLWM5F).
\end{acknowledgments}

\section*{Data Availability}

The data that support the findings of this study are available from the corresponding author upon reasonable request.

%


\end{document}